\documentclass[reprint,superscriptaddress,amsmath,amssymb,aps,prl]{revtex4-1}

\usepackage{graphicx}
\usepackage{hyperref}
\usepackage[english]{babel}

\begin{document}

\title{Dipole and quadrupole moments of \texorpdfstring{$^{73-78}$}{73-78}Cu as a test of the robustness of the \texorpdfstring{$Z=28$}{Z=28} shell closure near \texorpdfstring{$^{78}$Ni}{78Ni}}
    \author{R.~P.~de Groote}
    \email{ruben.degroote@kuleuven.be}
    \affiliation{KU Leuven, Instituut voor Kern- en Stralingsfysica, B-3001 Leuven, Belgium}
    \author{J.~Billowes}
    \affiliation{School of Physics and Astronomy, The University of Manchester, Manchester M13 9PL, UK}
    \author{C.~L.~Binnersley}
    \affiliation{School of Physics and Astronomy, The University of Manchester, Manchester M13 9PL, UK}
    \author{M.~L.~Bissell}
    \affiliation{School of Physics and Astronomy, The University of Manchester, Manchester M13 9PL, UK}
    \author{T.~E.~Cocolios}
    \affiliation{KU Leuven, Instituut voor Kern- en Stralingsfysica, B-3001 Leuven, Belgium}
    \author{T. Day Goodacre}
    \affiliation{Engineering Department, CERN, CH-1211 Geneva 23, Switzerland}
    \author{G.~J.~Farooq-Smith}
    \affiliation{KU Leuven, Instituut voor Kern- en Stralingsfysica, B-3001 Leuven, Belgium}
    \author{D.~V.~Fedorov}
    \affiliation{Petersburg Nuclear Physics Institute, 188350 Gatchina, Russia}
    \author{K.~T.~Flanagan}
    \affiliation{School of Physics and Astronomy, The University of Manchester, Manchester M13 9PL, UK}
    \author{S.~Franchoo}
    \affiliation{Institut de Physique Nucl\'{e}aire d'Orsay, F-91406 Orsay, France}
    \author{R.~F.~Garcia Ruiz}
    \affiliation{School of Physics and Astronomy, The University of Manchester, Manchester M13 9PL, UK}
    \author{\'{A}.~Koszor\'{u}s}
    \affiliation{KU Leuven, Instituut voor Kern- en Stralingsfysica, B-3001 Leuven, Belgium}
    \author{K.~M.~Lynch}
    \affiliation{Physics Department, CERN, CH-1211 Geneva 23, Switzerland}
    \author{G.~Neyens}
    \affiliation{KU Leuven, Instituut voor Kern- en Stralingsfysica, B-3001 Leuven, Belgium}
    \affiliation{Physics Department, CERN, CH-1211 Geneva 23, Switzerland}
    \author{F.~Nowacki}
    \affiliation{Institute Pluridisciplinaire Hubert Curien, 23 rue du Loess, F-67037 Strasbourg Cedex 2, France}
    \author{T.~Otsuka}
    \affiliation{KU Leuven, Instituut voor Kern- en Stralingsfysica, B-3001 Leuven, Belgium}
    \affiliation{Department of Physics, University of Tokyo, Hongo, Bunkyo-ku, Tokyo 113-0033, Japan}
    \affiliation{RIKEN Nishina Center, 2-1 Hirosawa, Wako, Saitama 351-0198, Japan}
    \affiliation{Center for Nuclear Study, University of Tokyo, Hongo, Bunkyo-ku, Tokyo 113-0033, Japan}
    \author{S.~Rothe}
    \affiliation{School of Physics and Astronomy, The University of Manchester, Manchester M13 9PL, UK}
    \affiliation{Engineering Department, CERN, CH-1211 Geneva 23, Switzerland}
    \author{H.~H.~Stroke}
    \affiliation{Department of Physics, New York University, New York, New York 10003, USA}
    \author{Y.~Tsunoda}
    \affiliation{Center for Nuclear Study, University of Tokyo, Hongo, Bunkyo-ku, Tokyo 113-0033, Japan}
    \author{A.~R.~Vernon}
    \affiliation{School of Physics and Astronomy, The University of Manchester, Manchester M13 9PL, UK}
    \author{K.~D.~A.~Wendt}
    \affiliation{Institut f\"{u}r Physik, Johannes Gutenberg-Universit\"{a}t, D-55128 Mainz, Germany}
    \author{S.~G.~Wilkins}
    \affiliation{School of Physics and Astronomy, The University of Manchester, Manchester M13 9PL, UK}
    \author{Z.~Y.~Xu}
    \affiliation{KU Leuven, Instituut voor Kern- en Stralingsfysica, B-3001 Leuven, Belgium}
    \author{X.~F.~Yang}
    \affiliation{KU Leuven, Instituut voor Kern- en Stralingsfysica, B-3001 Leuven, Belgium}

\begin{abstract}
Nuclear spins and precise values of the magnetic dipole and electric quadrupole moments of the ground-states of neutron-rich $^{76-78}$Cu isotopes were measured using the Collinear Resonance Ionization Spectroscopy (CRIS) experiment at ISOLDE, CERN. The nuclear moments of the less exotic $^{73,75}$Cu isotopes were re-measured with similar precision, yielding values that are consistent with earlier measurements. The moments of the odd-odd isotopes, and $^{78}_{29}$Cu ($N=49$) in particular, are used to investigate excitations of the assumed doubly-magic $^{78}$Ni core through comparisons with large-scale shell-model calculations. Despite the narrowing of the $Z=28$ shell gap between $N\sim45$ and $N=50$, the magicity of $Z=28$ and $N=50$ is restored towards $^{78}$Ni. This is due to weakened dynamical correlations, as clearly probed by the present moment measurements. 
\end{abstract}

\maketitle

The atomic nucleus is a complex system consisting of many interacting particles that nevertheless exhibits shell structure. Theoretical understanding of the shell gaps and, in particular, the magic nucleon numbers of 28 and 50, required the introduction of a strong spin-orbit force \cite{Mayer1949,Haxel1949}. In recent years, evidence interpreted as an evolution of the magic numbers when going away from the valley of stability has steadily grown, making the study of shell evolution one of the challenges in nuclear physics \cite{Otsuka2001,Sorlin2008}. The very exotic $^{78}$Ni ($Z=28, N=50$) nucleus is a cornerstone in this investigation, and a large volume of theoretical and experimental work has already been devoted to testing these magic numbers \cite{Otsuka2005,Sorlin2008,Hakala2008,Otsuka2010,Korgul2012,Marchi2014,Xu2014,Tsunoda2014,Coraggio2014,Sieja2012,Porquet2012,Hagen2016,Nowacki2016}. However, low production rates of isotopes in the vicinity of $^{78}$Ni (and, in particular, low production rates of the nickel isotopes themselves) have slowed experimental progress. 

Information on the size of the shell gaps as well as the correlations in the $^{78}$Ni system are required in order to confirm or refute its magicity. Nuclear dipole and quadrupole moments provide some of the most sensitive and direct tests for calculated nuclear wavefunctions \cite{Neyens2003}, and are therefore crucial observables to help settle the ongoing debate. Previously measured magnetic moments of copper ($Z=29$)  isotopes \cite{Cocolios2009,Cocolios2010,Flanagan2009,Vingerhoets2010,Vingerhoets2011} revealed the presence of proton excitations across $Z=28$, and therefore provided evidence for a breaking of this shell closure around $N=45$ \cite{Sieja2010}. Recent shell-model calculations predicted the size of the shell gap at $N=50$ to be smaller than at $N=45$ \cite{Sahin2017,Welker2017}. The reduction in the size of the shell gap could lead to a destabilization of $^{78}$Ni. However, any discussion of the magicity of $^{78}$Ni is incomplete without experimental information on proton-neutron excitations of the $^{78}$Ni core. As far as the copper isotopes are concerned, only the magnetic dipole moment of $^{77}$Cu has been measured beyond $N=46$ \cite{Koster2011}. Much-needed studies closer to $N=50$ were initiated through laser spectroscopy of gallium $(Z=31)$ \cite{Cheal2012,Procter2012b} and zinc $(Z=30)$ \cite{Yang2016,Wraith2017}. These results confirm the rapid evolution of proton and neutron shell gaps due to the strong $\pi f_{5/2}-\nu g_{9/2}$ interaction \cite{Otsuka2010}. Unfortunately, the presence of additional valence protons prevent these studies from pinning down the role of the nickel core excitations. Therefore, the need for experimental data on isotopes in close proximity of $^{78}$Ni remains, especially in light of the rapid shell evolution in this region of the nuclear chart.

\begin{table*}
\centering
\caption{Hyperfine parameters, magnetic dipole moments $\mu$ and electric quadrupole moments $Q$ obtained in this work, compared to literature \cite{Flanagan2009,Vingerhoets2010,Koster2011} and results from shell-model calculations using A3DA-m and PFSDG-U.}\label{tab:results}
\begin{tabular}{c|c|c|cccc|c|cccc}
    \toprule
    Mass & I & $A_l$\,(MHz) & \multicolumn{4}{c|}{$\mu$\,($\mu_N$)} & $B_u$\,(MHz)  & \multicolumn{4}{c}{$Q$\,(e\,fm$^2$)} \\
    \hline
    & & This work & This work &  Literature & A3DA-m & PFSDG-U & This work & This work &  Literature & A3DA-m & PFSDG-U \\
    \hline
    73 & $3/2^-$   & +4598(2)   & +1.7425(9)   &  +1.7426(8) &  +1.75 & -     &  -41(4)    & -23(2)  &  -20.0(8) & -18.15 & -      \\
    74 & $2^-$     & -2109(2)   & -1.0658(12)  &   +1.068(3) &  -1.12 & -1.04 &  +45(3)    & +25(3)  &  +26(3)   & +21.84 & +23.03 \\
    75 & $5/2^-$   &+1592.4(16) & +1.0058(10)  & +1.0062(13) &  +1.01 & +0.90 & -49(4)     &-27.5(17)& -26.9(16) & -27.66 & -29.20 \\
    76 & $3^-$     & -1437(2)   & -1.0895(15)  & -           &  -1.09 & -1.04 &  +62(4)    & +34(2)  &       -   & +33.73 & +35.86 \\
    77 & $5/2^-$   & +2524(3)   & +1.5945(17)  &   +1.61(5)  &  +1.62 & +1.55 & -47(4)     & -26(3)  &     -     & -26.44 & -28.10 \\
    78 & $ (6^-)$  &  +157(2)   & +0.238(3)    & 0.0(4)      &  -0.04 & +0.35 & +3(20)     & +2(10)  & -         & +13.40 & +10.57 \\
    79 & $(5/2^-)$ &  -         &    -         &    -        &  +1.96 & +1.96 & -          & -       &   -       & -22.71 & -22.50 \\
    \hline
\end{tabular}
\end{table*}

In this Rapid Communication we present precision measurements of the nuclear magnetic dipole and electric quadrupole moments of $^{76-78}$Cu ($Z=29$). Through comparisons of the experimental moments to state-of-the-art shell-model calculations, the evolution of the robustness of the $Z=28$ closure towards $N=50$ is directly investigated. The odd-odd isotopes, and in particular $^{78}$Cu, are advantageous probes for the structure of the $^{78}$Ni core. This advantage lies in the proximity of these isotopes to $^{78}$Ni, and in the sensitivity of their nuclear moments to simultaneously proton- and neutron excitations. This Rapid Communication therefore also highlights the importance of studying the static moments of odd-odd systems near doubly-closed shells for stringent testing of theoretical models.

\begin{figure}
    \includegraphics[width = 0.9\columnwidth]{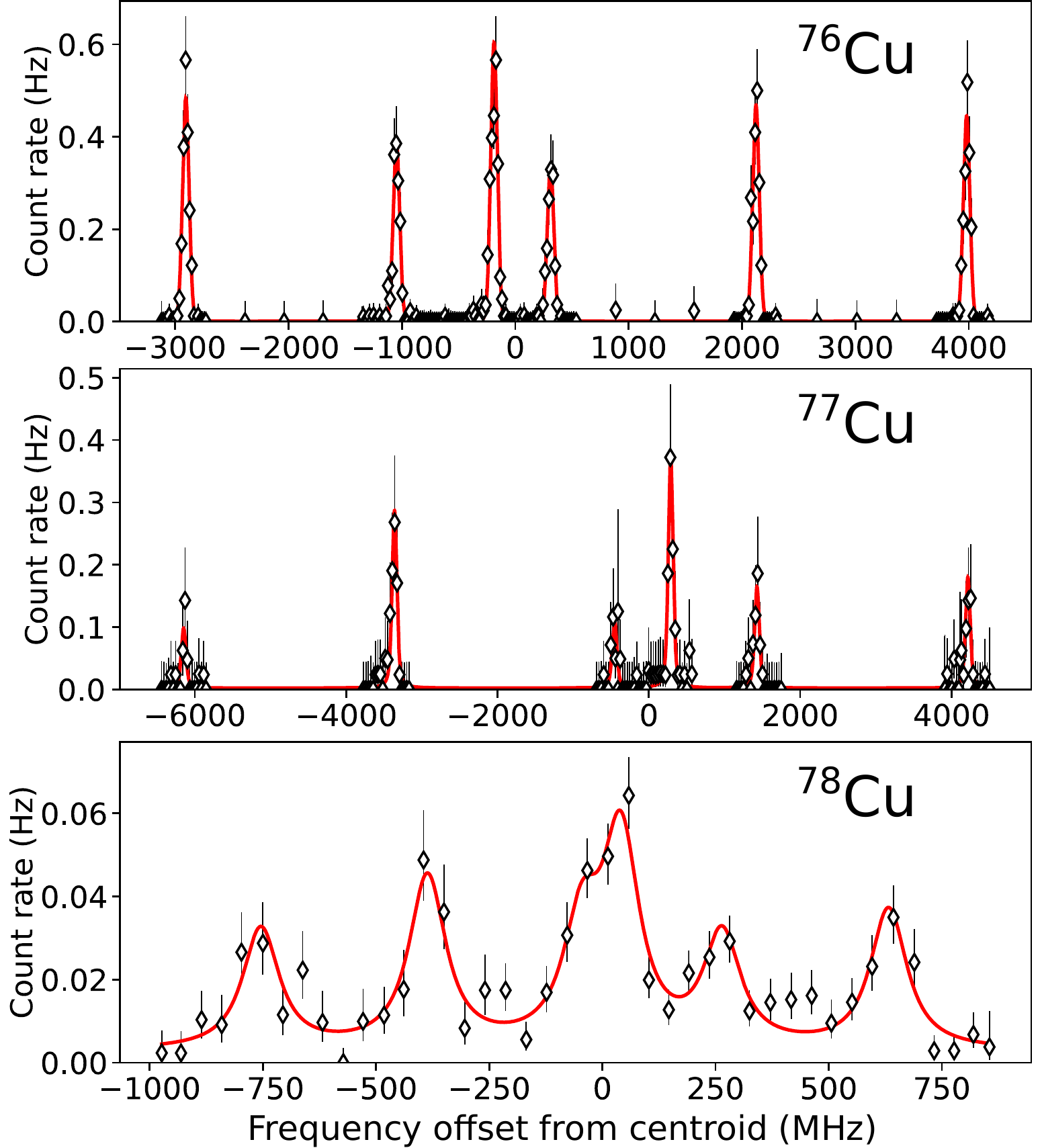}
    \caption{Example spectra of $^{76-78}$Cu plotted relative to their individual fitted centroid frequencies. A FWHM of 75\,MHz was achieved, due in part to the laser linewidth of $\approx$40\,MHz.}\label{fig:spectra}
\end{figure}

The copper isotopes of interest were produced at the ISOLDE facility at CERN by impinging 1.4\,GeV protons onto the neutron converter \cite{Koster2002} of a UC$_\text{x}$ target. After diffusing out of the target, the copper isotopes were laser ionized using the RILIS laser ion source \cite{Weissman2002,Koster2011,Rothe2016}, extracted at 30\,keV and mass separated by the high-resolution separator. The ions were then injected into the gas-filled radio-frequency linear Paul trap ISCOOL \cite{Mane2009,Franberg2008}. After 10 ms of accumulation in ISCOOL, the bunches of copper ions were extracted at 30\,keV. The electromagnetic moments of the copper isotopes were then measured using the Collinear Resonance Ionization Spectroscopy (CRIS) technique \cite{Flanagan2013,deGroote2015}, which combines high sensitivity with high resolution, thus allowing experiments on exotic beams with intensities as low as 20 ions/s (such as $^{78}$Cu). Upon entering the CRIS beamline, the ion bunches produced by ISCOOL were neutralized
by passing through a potassium-vapour charge-exchange cell \cite{Procter2012}. The remaining ionic fraction was electrostatically deflected, so only neutral atoms entered the laser-atom interaction region. These neutral atoms were collinearly overlapped with two pulsed laser beams to induce two-step resonant laser ionization. The ions of interest were deflected from the background neutral atoms, onto a micro-channel plate detector. This provides nearly background free detection: in this experiment, the suppression of collisional background counts was estimated at 1:10$^7$ at mass 78, representing an improvement by two orders of magnitude compared to our previous work \cite{Flanagan2013}. By recording the number of resonant ions as a function of the laser frequency, the hyperfine structure of the copper atoms was measured. The resonant ion detection rate for the most exotic isotope $^{78}$Cu was at most approximately 0.05/s. Only due to the high background suppression and efficiency could the hyperfine spectrum shown in Fig.\,\ref{fig:spectra} be measured in just 8 hours.

Atoms were resonantly excited from the $3d^{10}4s \ ^2S_{1/2}$ ground state to the $3d^9(^2D)4s4p(^3P^\circ) \ ^4P^\circ_{3/2}$ level at 40114.01\,cm$^{-1}$. This was achieved using an injection-locked titanium-sapphire laser system developed by the Johannes Gutenberg-Universit\"{a}t Mainz and the University of Jyv\"{a}skyl\"{a} \cite{Kessler2008,Sonnenschein2017}. Through pulsed amplification of 749\,nm continuous-wave seed light produced by a M-squared SolsTis titanium-sapphire laser, narrowband pulsed laser light was produced at a repetition rate of 1\,kHz. This laser light was then frequency tripled using two nonlinear crystals to produce the required narrowband ($\approx$40\,MHz) 249\,nm light for the resonant excitation step. Resonant ionization of these excited atoms was achieved using a 314.2444\,nm transition to the $3d^94s(^3D)4d \ ^4P_{3/2}$ auto-ionizing state at 71927.28\,cm$^{-1}$, using light produced by a frequency-doubled Spectron Spectrolase 4000 pulsed-dye laser pumped by a Litron LPY 601 50-100 PIV Nd:YAG laser, at a repetition rate of 100\,Hz. The 344(20)\,ns lifetime \cite{Kono1982} of the excited state allowed the 314\,nm laser pulse to be delayed by 50\,ns with respect to the first laser pulse, removing lineshape distortions and power broadening effects \cite{deGroote2015,deGroote2017} in the observed hyperfine spectra, without measurable losses in efficiency.

Table\,\ref{tab:results} shows the extracted hyperfine $A$ and $B$ parameters, defined as $A = \frac{\mu B_0}{IJ} \quad B = e Q V_{zz}$, with $B_0$ and $V_{zz}$ respectively the magnetic field and the electric field gradient generated by the electrons at the nucleus. By combining these equations with literature values, the dipole and quadrupole moments of $^{73-78}$Cu were obtained, listed in Table\,\ref{tab:results}. For every scan, the hyperfine parameters and uncertainty estimates were extracted using a maximum-likelihood optimization, performed using the SATLAS code \cite{Gins2017}. A Voigt lineshape was used, with additional modifications to model small lineshape asymmetries (to be detailed in a later publication). Final values for each isotope were then computed as a dispersion-corrected weighted mean, using the inverse square of the statistical fit uncertainties as weights. All reported uncertainties represent the 68\,\% credible intervals. The moments were calculated with $^{65}$Cu as the reference, using $\mu = 2.3817(3)\mu_{\text{N}}$ \cite{Lutz1978}, $Q = -19.5(4)\,\text{e\,fm}^2$ \cite{Sternheimer1972,stone2005}, $A=6284.389972(60)$\,MHz \cite{Figger1967} and $B=-35.0(2)$\,MHz \cite{Bucka1967}. The results are compared to literature values where available, showing excellent agreement and similar precision, demonstrating the performance of this CRIS set-up.

The ratio of the hyperfine $A$ parameters of the atomic ground and excited state is a constant for all isotopes of an element (aside from the negligibly small hyperfine anomaly \cite{Bucka1967,Locher1974}). This was verified with the data obtained on $^{63-75}$Cu, as will be discussed in a later publication. If this ratio can be determined with sufficient precision, it provides a model-independent way to determine the nuclear spin. Based on this ratio, firm spin assignments of $I=3$ for $^{76}$Cu and $I=5/2$ for $^{77}$Cu can be made, consistent with earlier tentative values of 3 or 4 for $^{76}$Cu \cite{VanRoosbroeck2005} and $5/2$ for $^{77}$Cu \cite{Patronis2009,Ilyushkin2009,Koster2011}. The larger uncertainties and the reduced sensitivity of the $A$-ratio for higher values of $I$ prevents a firm spin assignment for $^{78}$Cu using this test. However, shell-model calculations (which will be introduced later) were found to be in best agreement for spin 6. This value provides further evidence for the earlier tentative assignment presented in \cite{gross2009}. For the remainder of this Rapid Communication, a spin 6 will therefore be assumed for $^{78}$Cu. Note that, while the agreement is best for $I=6$, the other spin options $I=4,5,7$ also result in fair agreement. More details on these spin assignments will be reported in a later publication.

\begin{figure}[ht!]
\centering
    \includegraphics[width = 0.78\columnwidth]{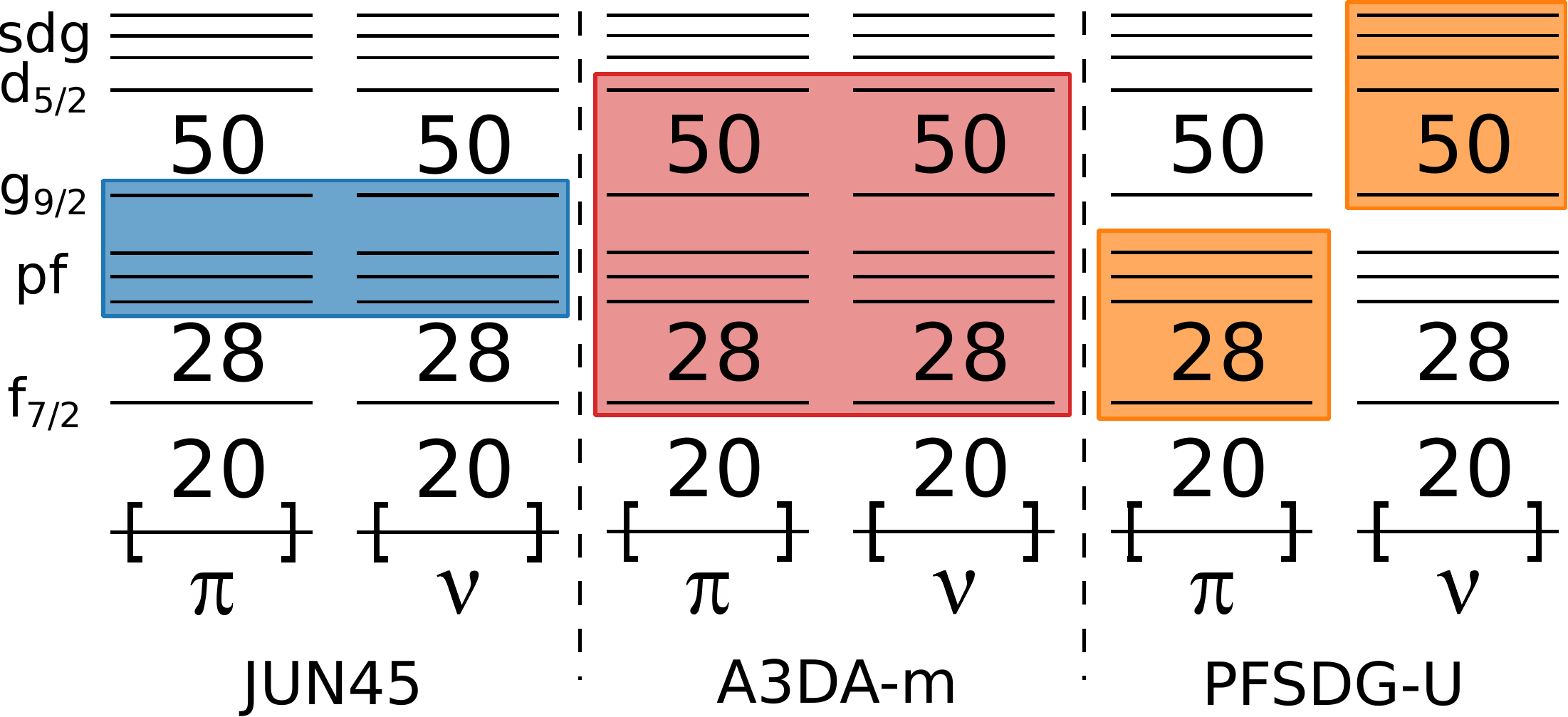}
    \caption{Model spaces used in this Rapid Communication for the JUN45, A3DA-m and PFSDG-U calculations.}
    \label{fig:model_spaces}
\end{figure}

An overview of the model spaces used in the shell-model calculations presented in this Rapid Communication is shown in Fig.\,\ref{fig:model_spaces}. Calculations starting from a $^{56}$Ni core, with protons and neutrons restricted to the $f_{5/2},p_{3/2},p_{1/2},g_{9/2}$ shell between $N,Z=28$ and $N,Z=50$ are performed using the JUN45 interaction \cite{Honma2009}. Calculations in a more extended space which includes the $f_{7/2}$ orbits below 28 and the $d_{5/2}$ orbits above 50 were performed using the A3DA-m \cite{Shimizu2012,Tsunoda2014} interaction. The results for the most neutron-rich isotopes, from $N=45$ onwards, are also compared to calculations including the full negative parity $pf$ shell for protons and full positive parity $sdg$ shell for neutrons using the recently developed PFSDG-U interaction \cite{Nowacki2016}. For all magnetic moment calculations, a spin-quenching factor of 0.75 was used, based on the value of 0.75$\pm 0.02$ suggested in \cite{Caurier2005} for pf-shell nuclei. Effective orbital g-factors were taken to be $1.1$ and $-0.1$ for protons and neutrons respectively. For the quadrupole moment calculations, effective charges of $e_p = 1.5, e_n = 1.1$ were assumed for JUN45 \cite{Honma2009}. For A3DA-m and PFSDG-U calculations the microscopically derived effective charges of $e_p = 1.31, e_n = 0.46$ \cite{Dufour1996} were chosen. Table \ref{tab:results} summarizes the results of these calculations. 

\begin{figure}[ht!]
    \centering
    \includegraphics[width=\columnwidth]{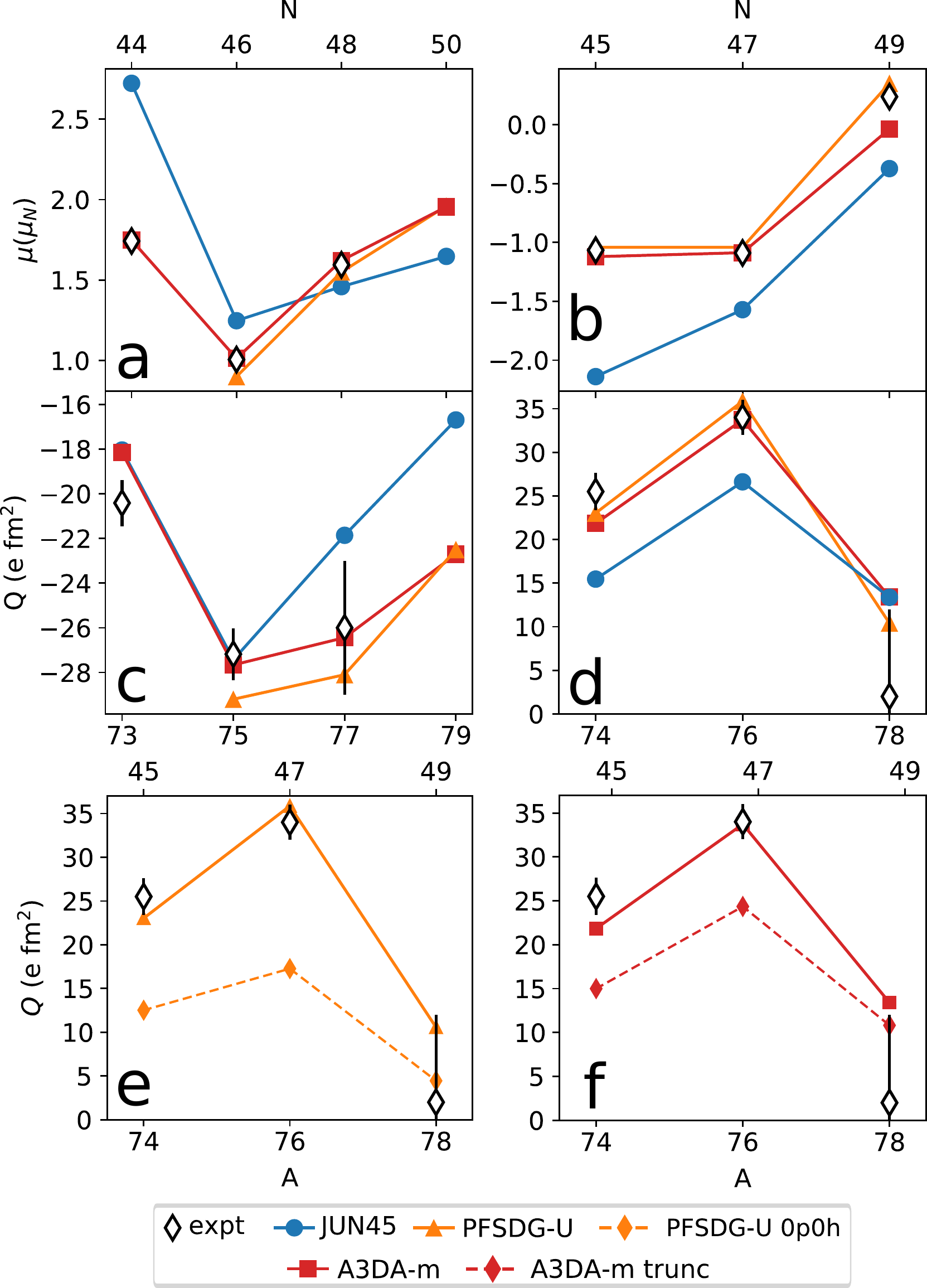}
    \caption{Panel a (b): Dipole moments of the odd-even (odd-odd) copper isotopes. Panel c (d): Quadrupole moments of the odd-even (odd-odd) copper isotopes. Panels e and f: Quadrupole moments of the odd-odd copper isotopes compared to PFSDG-U calculations without cross-shell excitations (labeled `0p0h') and A3DA-m calculations where the $d_{5/2}$ orbit is removed from the model space (labeled `trunc').} 
    \label{fig:all}
\end{figure}

In panels a and c of Fig.\,\ref{fig:all}, the experimental and theoretical moments of the odd-$A$ copper isotopes between $N=44$ and $N=50$ are presented. Where possible, the weighted mean of this work and literature values are plotted. Earlier calculations for the odd-even isotopes, which were performed in a model space that included the full $pf$-shell for protons and limited the neutrons to orbits up to $N=50$, illustrated the need for excitations of protons across $Z=28$ \cite{Sieja2010}. Excitations of neutrons across $N=50$, absent due to the model space that was used, seemed unneeded, and quadrupole moments were furthermore not compared to. The importance of these cross-shell proton excitations is confirmed in panel a: the A3DA-m and PFSDG-U magnetic moments are in excellent agreement with the data for all odd-A copper isotopes, while the JUN45 values deviate, in particular for $^{73}$Cu. From the comparison of the quadrupole moments of the odd-A copper isotopes to the different calculations, shown in panel c, no firm conclusions can be made considering the importance of neutron excitations. This is due to the larger uncertainties on $Q(^{77}$Cu). 

However, the odd-odd moments are much more sensitive probes to test the validity of the different wave functions. The magnetic moment of the high-spin $^{78}$Cu in particular provides a clean probe for purity of the nuclear wavefunction. As can be seen in panels b and d in Fig.\,\ref{fig:all}, the JUN45 calculations do not agree with the values of both $\mu$ and $Q$ for the odd-odd isotopes. The A3DA-m and the PFSDG-U predictions do agree very well with experimental values of $\mu$ and $Q$, despite the increased complexity of shell-model calculations for such systems. This improved agreement can be attributed to the presence of excitations of protons as well as neutrons across the $Z=28$ and $N=50$ gaps respectively, which are absent in the JUN45 calculations.

The need for these core excitations is further illustrated in panel e of Fig\,\ref{fig:all}, which compares to odd-odd $Q$ moments from PFSDG-U calculations where all cross-shell proton and neutron excitations are blocked, and panel f where A3DA-m calculations with neutrons restricted to the $f_{7/2},f_{5/2},p_{3/2},p_{1/2}$ and $g_{9/2}$ orbits are shown. In these truncated A3DA-m calculations, excitations across the $N=50$ gap are therefore absent. Both restricted calculations yield moments that deviate from the experimental values, highlighting the sensitivity of the moments of these odd-odd copper isotopes to excitations of the underlying nickel core, and furthermore illustrate the importance of neutron excitations. For $^{78}$Cu, the truncated calculations deviate less from the full-space calculations, which illustrates the reduction of correlations. The moments of the odd-odd isotopes are therefore the key to uncovering the evolution of both proton and neutron correlations. 

\begin{figure}[ht!]
    \centering
    \includegraphics[width = \columnwidth]{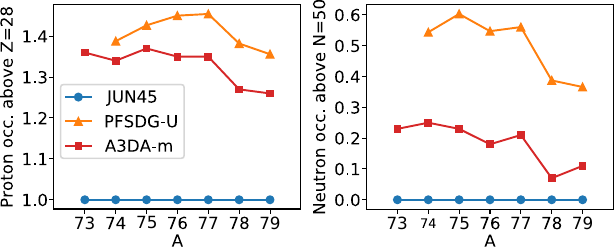}
    \caption{Comparison of the occupations above $Z=28$ and $N=50$ for the JUN45, A3DA-m and PFSDG-U calculations.}
    \label{fig:occupations}
\end{figure}

The persistence of proton- and neutron excitations across $Z=28$ and $N=50$ beyond $N=45$ can be investigated in more detail by looking at the occupations of the proton and neutron orbits above the $Z=28$ and the $N=50$ gaps, shown in Fig.\,\ref{fig:occupations}. These occupancies remain rather constant from $^{73}$Cu to $^{77}$Cu for both A3DA-m and PFSDG-U. However, proton and neutron cross-shell excitations decrease suddenly from $^{78}$Cu onwards. These weakened core excitations indicate a restoration of the $Z=28$ and $N=50$ shell closures when $N=50$ is approached, despite the reduction in the size of the single-particle $Z=28$ shell gap. This reduction is due to the central and tensor forces \cite{Otsuka2005,Otsuka2010}, and was recently calculated to be 2\,MeV between $N=40$ and $N=50$ \cite{Tsunoda2014,Sahin2017,Welker2017}. Together with the recently measured $^{77}$Cu level scheme \cite{Sahin2017} and the recent mass-measurements of copper isotopes up to $^{79}$Cu \cite{Welker2017}, clear evidence in favor of a doubly-magic $^{78}$Ni has therefore been obtained. This conclusion relies crucially on the experimental verification of the weakening correlations, which the current measurements of the moments of the odd-odd isotopes finally provide.

In conclusion, this Rapid Communication reports on the first high-resolution laser spectroscopy measurements of $^{76-78}$Cu. Since $^{78}$Cu was only produced at a rate of 20 particles per second, an improvement of the background suppression by two orders of magnitude as compared to previous experiments was required. The excellent agreement of the moments of the odd-odd copper isotopes with state-of-the-art shell-model calculations in extended model spaces provides a clear indication for the weakening of correlations towards $N=50$: both the proton and neutron excitations across $Z=28$ and $N=50$ decrease for $^{78,79}$Cu. Strong evidence in favour of a restoration of the magicity of $N=50$ and $Z=28$ is therefore obtained, despite the reduction in the size of the shell gaps. This investigation into the doubly magic character of $^{78}$Ni relies crucially on measurements of the moments of odd-odd nuclei. While their increased complexity can make calculation and interpretation more difficult, the unique information contained in their nuclear moments helps to clarify the subtle interplay of the single-particle properties and collective correlations in exotic nuclei, where experimental data are otherwise scarce. Finally, the nuclear spins of $^{76,77}$Cu were measured to be 3 and 5/2 respectively.

\begin{acknowledgments}
We acknowledge the support of the ISOLDE collaboration and technical teams, and the University of Jyv\"{a}skyl\"{a} for the use of the injection-locked cavity. The Monte Carlo Shell-Model calculations were performed on K computer at RIKEN AICS (hp160211, hp170230). This work was supported in part by FNPMLS ERC Consolidator Grant no.~648381, MEXT as ``Priority Issue on Post-K computer'' (Elucidation of the Fundamental Laws and Evolution of the Universe) and JICFuS, the BriX Research Program No.~P7/12, FWO-Vlaanderen (Belgium), GOA 15/010 from KU Leuven, the Science and Technology Facilities Council consolidated grant ST/F012071/1 and continuation grant ST/J000159/1, the EU Horizon2020 research and innovation programme through ENSAR2 (no. 654002), and Ernest Rutherford Grant No. ST/L002868/1. We acknowledge the financial aid of the Ed Schneiderman Fund at New York University.
\end{acknowledgments}

\bibliography{biblio}

\end{document}